\documentstyle[amssymb,amsmath]{article}
\begin{document}

\title{Critical phenomena and critical relations}
\author{M.Matlak\thanks{%
matlak@us.edu.pl} and B.Grabiec \\
%EndAName
Institute of Physics, Silesian University, 4 Uniwersytecka,\\
40-007 Katowice, Poland}
\maketitle

\begin{abstract}
We consider systems which exhibit typical critical dependence of the
specific heat: $\Delta c\varpropto (T_C-T)^{-\gamma }$ ($T<T_C$); $\Delta
c\varpropto (T-T_C)^{-\gamma ^{\prime }}$ ($T>T_C$) where $\gamma $,$\gamma
^{\prime }$ are critical exponents ($\gamma =\alpha $ for $\Delta c=\Delta
c_{p,N},$ $\gamma =\overline{\alpha }$ for $\Delta c=\Delta c_{V,N}$), as
well as, the case when $\Delta c\varpropto (\ln \mid T_C-T\mid )^a$ ($%
a=\frac 13$, uniaxial ferroelectrics; $a=1$, liquid $He^4$). Starting from
the critical behaviour of the specific heat we find the Gibbs (Helmholtz)
potential in the vicinity of the critical point for each case separately. We
derive in this way many exact critical relations in the limit $T\rightarrow
T_C$ which remain the same for each considered case. They define a new class
of universal critical relations independent from the underlying microscopic
mechanism and the symmetry of these systems. The derived relations are valid
for a very broad class of magnetic, ferroelectric and superconducting
materials, as well as, for liquid $He^4$.

\end{abstract}

%\address{Institute of Physics, Silesian University, 4 Uniwersytecka,\\
%40-007 Katowice, Poland}
%\submitted{}

\section{Introduction}

Critical phenomena in solids, connected with second order phase transitions,
belong to very attractive and intriguing problems in physics. Many
theoretical models and methods have been applied to resolve these
interesting questions (cf e.g. Refs [1]-[5] for a review and the original
papers cited therein), however, the most important, general and successfull
method, applied in this area of physics, seems to be still the
phenomenological approach, initiated by the pioneer works of Landau, Refs
[6], [7]. This approach has succsessfully been applied to superconducting
systems resulting in many useful critical relations, derived in Ref. [8].
For many other systems (magnetic, ferroelectric, liquid $He^4$, etc.) this
approach unfortunately fails because of critical exponents (cf e.g. Refs
[1]-[5]). We can, however, improve this approach and calculate the Gibbs
(Helmholtz) potential near the critical point from the observed specific
heat anomaly ($T<T_C;T>T_C$). Similarly to Ref. [8] we include into the
considerations the critical behaviour of the chemical potential as indicator
of phase transitions in a more general sense (see Ref. [8] and papers cited
therein). We derive in this way many interesting thermodynamical relations
(cross-relations) at the critical point in the limit $T\rightarrow T_C$.
These exact relations seem to be of fundamental importance for critical
phenomena in solids and should generally be valid for a very broad class of
magnetic or ferroelectric systems, as well as, for superconducting systems
as a special case. Surprisingly, they are also valid in the case of
logarithmic divergence of the specific heat (uniaxial ferroelectrics and
liquid $He^4$). The derived relations possess very similar form to the
critical relations presented in Ref. [8] and define a new class of exact
relations at the critical point independent from the underlying microscopic
mechanism leading to second order phase transitions and independent from the
symmetry of the system.

\section{Phenomenological Gibbs (Helmholtz) potentials and critical relations
}

Performing experimental measurements of the temperature dependence of the
specific heat $c_{p,N}$ at constant pressure $p$ and constant number of
particles $N$ we can find that for magnetic, ferroelectric or
superconducting systems $c_{p,N}$ exhibits a pronounced anomaly at the
critical point $T=T_C$ (cf e.g. Refs [1]-[5]). Independently from the
considered system (here: magnetic, ferroelectric or superconducting) it is
convenient to introduce a reference system which possesses exactly the same
structure (symmetry), atomic masses, etc., the specific heat of this system $%
c_{p,N}^{(0)}$ exhibits, however, no anomaly at $T=T_C$. Let us introduce an
auxiliary quantity
\begin{equation}
\Delta c_{p,N}(T)\equiv c_{p,N}(T)-c_{p,N}^{(0)}(T).
\end{equation}
A very broad class of materials (magnetic, ferroelectric or superconducting)
shows a general, critical behaviour (see e.g. [1]-[5]) of the type\footnote{%
The case of the logarithmic divergence of the specific heat will be
considered in the next Section.}
\begin{equation}
\Delta c_{p,N}(T)\varpropto \left\{
\begin{array}{ll}
(T_C-T)^{-\alpha }, & T<T_C \\
(T-T_C)^{-{\alpha }^{\prime }}, & T>T_C
\end{array}
\right.
\end{equation}
where $\alpha ,$ $\alpha ^{\prime }$ are critical exponents, depending on
the material ($\alpha ,$ $\alpha ^{\prime }>0)$. For magnetic and
ferroelectric systems $\Delta c_{p,N}$ usually diverges at the critical
point, for superconducting ones $\Delta c_{p,N}$ exhibits only a finite
jump. For simplicity, we restrict ourselves to the case $T<T_C$ and we
assume (see expression (2)) that in the vicinity of the critical point
\begin{equation}
\Delta c_{p,N}(T)=A(p,N)(T_C-T)^{-\alpha }
\end{equation}
where $A(p,N)$ is a parameter. The critical temperature $T_C$ should also be
considered as a function of $p$ and $N$ ($T_C=T_C(p,N)$), as suggested by
experimental results. The thermodynamics of the system in the vicinity of
the critical point ($T<T_C$) can be described by the Gibbs potential $%
Z(T,p,N)$. We can formally write
\begin{equation}
Z(T,p,N)={Z_0}(T,p,N)+\Delta Z
\end{equation}
where
\begin{equation}
\Delta Z={Z}-Z_0,
\end{equation}
$Z_{0\text{ }}$is the Gibbs potential of the reference system and $%
c_{p,N}^{(0)}=-T\left( \frac{\partial ^2Z_0}{\partial T^2}\right) _{p,N}$.
The analytical form of $\Delta Z$ should necessarily be chosen in such a way
that the thermodynamical relation
\begin{equation}
\Delta c_{p,N}=-T\left( \frac{\partial ^2(\Delta Z)}{\partial T^2}\right)
_{p,N}
\end{equation}
reproduces the critical behaviour of $\Delta c_{p,N}$, given by the formula
(3). It is easy to find that the leading term of $\Delta Z$ should have the
general form\footnote{%
It is possibly to find $\Delta Z$ exactly when integrate the expression (6)
with the use of the formula (3). The result is given in the Appendix A. To
avoid the presentation of very long expressions we restrict ourselves to
consider only the leading term in $\Delta Z$, given by the formula (7). This
approach (see Appendix A) leads to the same, exact critical relations
(18)-(22) as in the case when the complete expression for $\Delta Z$ (see
the formula (A.6)) is used.}
\begin{equation}
\Delta Z=D(p,N)(T_C-T)^{2-\alpha }
\end{equation}
where $D(p,N)$ is a parameter. Really, when applying (6) and (7) we obtain
\begin{eqnarray}
\Delta c_{p,N} &=&-T(2-\alpha )(1-\alpha )D(p,N)(T_C-T)^{-\alpha }  \nonumber
\\
&\thickapprox &-T_C(2-\alpha )(1-\alpha )D(p,N)(T_C-T)^{-\alpha }  \nonumber
\\
&=&A(p,N)(T_C-T)^{-\alpha }
\end{eqnarray}
where
\begin{equation}
A(p,N)=-T_C(2-\alpha )(1-\alpha )D(p,N).
\end{equation}
Thus, the asymptotic form of the Gibbs potential (4) in the vicinity of the
critical point ($T<T_C$) has to be equal to
\begin{equation}
Z(T,p,N)={Z_0}(T,p,N)+D(p,N)(T_C-T)^{2-\alpha }.
\end{equation}
The phenomenological relations (3) and (10), resulting from the
experimental, critical behaviour of a broad class of systems can easily be
used to derive many critical relations. The chemical potential $\mu =(\frac{%
\partial Z}{\partial N})_{T,p}$ ($\mu _0=(\frac{\partial Z_0}{\partial N}%
)_{T,p}$) and the volume of the system $V=(\frac{\partial Z}{\partial p}%
)_{T,N}$ ($V_0=(\frac{\partial Z_0}{\partial p})_{T,N}$) can easily be found
from (10). We obtain %\begin{widetext}
\begin{eqnarray}
\mu &=&\mu _0+\left( \frac{\partial D}{\partial N}\right)
_p(T_C-T)^{2-\alpha }  \nonumber \\
&+&(2-\alpha )D(T_C-T)^{1-\alpha }\left( \frac{\partial T_C}{\partial N}%
\right) _p
\end{eqnarray}
and
\begin{eqnarray}
V &=&V_0+\left( \frac{\partial D}{\partial p}\right) _N(T_C-T)^{2-\alpha }
\nonumber \\
&+&(2-\alpha )D(T_C-T)^{1-\alpha }\left( \frac{\partial T_C}{\partial p}%
\right) _N
\end{eqnarray}
%\end{widetext}
where we have taken into account that $T_C=T_C(p,N)$. With the use of (11)
and (12) we obtain %\begin{widetext}
\begin{eqnarray}
\Delta \left( \frac{\partial \mu }{\partial T}\right)
_{p,N}&=&-(T_C-T)^{-\alpha }(2-\alpha )\left[ {}\right. \left( \frac{\partial D%
}{\partial N}\right) _p(T_C-T) \nonumber \\
&&+(1-\alpha )D\cdot \left( \frac{\partial T_C}{\partial N}\right) _p\left. {}\right] ,
\end{eqnarray}

\begin{eqnarray}
\Delta \left( \frac{\partial V}{\partial T}\right) _{p,N}&=&-(T_C-T)^{-\alpha
}(2-\alpha )\left[ {}\right. \left( \frac{\partial D}{\partial p}\right)
_N(T_C-T) \nonumber \\
&&+(1-\alpha )D\cdot \left( \frac{\partial T_C}{\partial p}\right)
_N\left. {}\right] ,
\end{eqnarray}

\begin{eqnarray}
\Delta \left( \frac{\partial \mu }{\partial p}\right) _{T,N}
&=&(T_C-T)^{-\alpha }\Big\{ \frac{\partial ^2D}{\partial p\partial N}%
(T_C-T)^2  \nonumber \\
&&+(2-\alpha )(T_C-T)\Big[ \left( \frac{\partial D}{\partial N}\right)
_p\left( \frac{\partial T_C}{\partial p}\right) _N  \nonumber \\
&&+\left( \frac{\partial D}{\partial p}\right) _N\left( \frac{\partial T_C}{%
\partial N}\right) _p +D\cdot \left( \frac{\partial ^2T_C}{\partial
p\partial N}\right) \Big]  \nonumber \\
&&+(2-\alpha )(1-\alpha )D\cdot \left( \frac{\partial T_C}{\partial p}%
\right) _N\left( \frac{\partial T_C}{\partial N}\right) _p\Big\},
\end{eqnarray}

\begin{eqnarray}
\Delta \left( \frac{\partial \mu }{\partial N}\right) _{T,p}
&=&(T_C-T)^{-\alpha } \Big\{ \left( \frac{\partial ^2D}{\partial N^2}
\right)_{p} (T_{C}-T)^2  \nonumber \\
&&+(2-\alpha )(T_C-T)\Big[ 2\left( \frac{\partial D}{\partial N}\right)
_p\left( \frac{\partial T_C}{\partial N}\right) _p +D\cdot \left( \frac{%
\partial ^2T_C}{\partial N^2}\right) _p\Big]  \nonumber \\
&&+(2-\alpha )(1-\alpha )D\cdot \Big[ \left( \frac{\partial T_C}{\partial N}%
\right) _p\Big]^2\Big\}
\end{eqnarray}
and
\begin{eqnarray}
\Delta \left( \frac{\partial V}{\partial p}\right) _{T,N}
&=&(T_C-T)^{-\alpha }\Big\{ \left( \frac{\partial ^2D}{\partial p^2}\right)
_N(T_C-T)^2  \nonumber \\
&&+(2-\alpha )(T_C-T)\Big[ 2\left( \frac{\partial D}{\partial p}\right)
_N\left( \frac{\partial T_C}{\partial p}\right) _N
+D\cdot \left( \frac{\partial ^2T_C}{\partial p^2}\right) _N\Big]  \nonumber \\
&&+(2-\alpha )(1-\alpha )D\cdot \Big[ \left( \frac{\partial T_C}{\partial p}%
\right) _N\Big]^2\Big\}
\end{eqnarray}
where $\Delta X\equiv X(T)-X_0(T)$ has exactly the same interpretation as in
the formula (1). %\end{widetext}
It is interesting to note that for magnetic (ferroelectric) systems $\Delta
c_{p,N}$, $\Delta (\frac{\partial \mu }{\partial T})_{p,N}$, $\Delta (\frac{%
\partial V}{\partial T})_{p,N}$, $\Delta (\frac{\partial \mu }{\partial p}%
)_{T,N}$, $\Delta (\frac{\partial \mu }{\partial N})_{T,p}$ and $\Delta (%
\frac{\partial V}{\partial p})_{T,N}$ , generally, diverge at the critical
point. However, when e.g. divide (13)-(17) by (3) the ''dangerous'' term $%
(T_C-T)^{-\alpha }$ cancels out (see also (9)) and taking the limit $%
T\rightarrow T_C^{-}$ we obtain the following relations 
\begin{equation}
\lim\limits_{T\rightarrow T_C^{-}}\frac{\Delta (\frac{\partial \mu }{%
\partial T})_{p,N}}{\Delta c_{p,N}}=\left( \frac{\partial \ln T_C}{\partial N%
}\right) _p,
\end{equation}

\begin{equation}
\lim\limits_{T \rightarrow T_C^-} \frac{\Delta (\frac{\partial V}{\partial T}%
)_{p,N}}{\Delta c_{p,N}}=\left( \frac{\partial \ln T_C}{\partial p}
\right)_N,
\end{equation}

\begin{equation}
\lim\limits_{T \rightarrow T_C^-} \frac{\Delta (\frac{\partial \mu}{\partial
p})_{T,N}}{\Delta c_{p,N}}=-\left( \frac{\partial \ln T_C}{\partial N}
\right)_p \left( \frac{\partial T_C}{\partial p} \right)_N,
\end{equation}

\begin{equation}
\lim\limits_{T\rightarrow T_C^{-}}\frac{\Delta (\frac{\partial \mu }{%
\partial N})_{T,p}}{\Delta c_{p,N}}=-\left( \frac{\partial \ln T_C}{\partial
N}\right) _p\left( \frac{\partial T_C}{\partial N}\right) _p
\end{equation}
and 
\begin{equation}
\lim\limits_{T\rightarrow T_C^{-}}\frac{\Delta (\frac{\partial V}{\partial p}%
)_{T,N}}{\Delta c_{p,N}}=-\left( \frac{\partial \ln T_C}{\partial p}\right)
_N\left( \frac{\partial T_C}{\partial p}\right) _N.
\end{equation}
It, however, means that the limits exist and the derived relations are exact%
\footnote{%
The additional terms to $\Delta Z$ (see Appendix A) modify only the
expressions (11)-(17). By forming the quotients (18)-(22) in the limit $%
T\rightarrow T_C^{-}$ their contribution is equal to zero (see Appendix A)
and therefore the critical relations (18)-(22) are, in fact, exact.}. For
classic superconductors (${\alpha }=0$) all the quantities (3) and (13)-(17)
are finite and therefore we can drop the $\lim\limits_{T\rightarrow T_C^{-}}$
in the expressions (18)-(22) and we obtain the formulae (29)-(33) from Ref.
[8]. In orther words the formulae (18)-(22) are valid for a broad class of
magnetic, ferroelectric and superconducting systems. Because of Maxwell
relations $\left( \frac{\partial S}{\partial p}\right) _{T,N}=-\left( \frac{%
\partial V}{\partial T}\right) _{p,N}$, $\left( \frac{\partial S}{\partial N}%
\right) _{T,p}=-\left( \frac{\partial \mu }{\partial T}\right) _{p,N}$ and $%
\left( \frac{\partial V}{\partial N}\right) _{T,p}=\left( \frac{\partial \mu 
}{\partial p}\right) _{T,N}$ there are only five independent relations of
the type (13)-(17). Because $\Delta \left( \frac{\partial S}{\partial p}%
\right) _{T,N}=-\Delta \left( \frac{\partial V}{\partial T}\right) _{p,N}$, $%
\Delta \left( \frac{\partial S}{\partial N}\right) _{T,p}=-\Delta \left( 
\frac{\partial \mu }{\partial T}\right) _{p,N}$ and $\Delta \left( \frac{%
\partial V}{\partial N}\right) _{T,p}=\Delta \left( \frac{\partial \mu }{%
\partial p}\right) _{T,N}$ the corresponding critical relation with the
presence of $\Delta \left( \frac{\partial S}{\partial p}\right) _{T,N}$, $%
\Delta \left( \frac{\partial S}{\partial N}\right) _{T,p}$ and $\Delta
\left( \frac{\partial V}{\partial N}\right) _{T,p}$can easily be obtained
from (19),(18) and (20), respectively. Besides, it is very easy to obtain
many other relations when e.g. divide (14) by (13) and take the limit 
\begin{equation}
\lim\limits_{T\rightarrow T_C^{-}}\frac{\Delta (\frac{\partial V}{\partial T}%
)_{p,N}}{\Delta (\frac{\partial \mu }{\partial T})_{p,N}}=\frac{(\frac{%
\partial T_C}{\partial p})_N}{(\frac{\partial T_C}{\partial N})_p}
\end{equation}
and so on. Therefore there is no need to write all of them explicitely here.
It is also possible to calculate the derivatives $(\frac{\partial T_C}{%
\partial N})_p$ , $(\frac{\partial T_C}{\partial p})_N$ from the relations
(18), (19) and isert them into (20)-(22). In such a way we obtain
cross-relations %\begin{widetext}
\begin{equation}
\lim\limits_{T\rightarrow T_C^{-}}\frac{\Delta \left( \frac{\partial \mu }{%
\partial p}\right) _{T,N}}{\Delta c_{p,N}}=-T_C\big[\lim\limits_{T%
\rightarrow T_C^{-}}\frac{\Delta \left( \frac{\partial \mu }{\partial T}%
\right) _{p,N}}{\Delta c_{p,N}}\big] \cdot \big[ \lim\limits_{T\rightarrow
T_C^{-}}\frac{\Delta \left( \frac{\partial V}{\partial T}\right) _{p,N}}{%
\Delta c_{p,N}}\big]
\end{equation}

\begin{equation}
\lim\limits_{T \rightarrow T_C^-} \frac{\Delta \left( \frac{\partial \mu }{%
\partial N}\right)_{T,p}}{\Delta c_{p,N}}= -T_{C}\big[\lim\limits_{T
\rightarrow T_C^-} \frac{\Delta \left( \frac{\partial \mu }{\partial T}%
\right)_{p,N}}{\Delta c_{p,N}} \big]^2
\end{equation}
and 
\begin{equation}
\lim\limits_{T\rightarrow T_C^{-}}\frac{\Delta \left( \frac{\partial V}{%
\partial p}\right) _{T,N}}{\Delta c_{p,N}}=-T_C\big[\lim\limits_{T%
\rightarrow T_C^{-}}\frac{\Delta \left( \frac{\partial V}{\partial T}\right)
_{p,N}}{\Delta c_{p,N}}\big]^2.
\end{equation}
%\end{widetext}
In the case of superconducting systems we can drop the $\lim\limits_{T%
\rightarrow T_C^{-}}$ and the formulae (24)-(26) coincide with the formulae
(34)-(36) from Ref. [8]. There are, however, again many other possibilities
to form similar cross-relations which can be found with ease when starting
from the formulae (3), (9) and (13)-(17).

We can also derive additional relations for systems at constant volume $V$.
Similarly to (2) we can write (cf e.g. [1]-[5]) 
\begin{equation}
\Delta c_{V,N}(T)\varpropto \left\{ 
\begin{array}{ll}
(T_C-T)^{-\overline{\alpha }}, & T<T_C \\ 
(T-T_C)^{-\overline{\alpha ^{\prime }}}, & T>T_C
\end{array}
\right.
\end{equation}
where $\overline{\alpha }$ and $\overline{\alpha ^{\prime }}$ are critical
exponents. We restrict ourselves again to the vicinity of the critical point
for $T<T_C$ and we can assume that 
\begin{equation}
\Delta c_{V,N}=\overline{A}(V,N)(T_C-T)^{-\overline{\alpha }}
\end{equation}
where $\overline{A}(V,N)$ is a parameter. The corresponding leading term of
the Helmholtz free energy has to be in the form (cf also (10)) 
\begin{equation}
F(T,V,N)=F_0(T,V,N)+\overline{D}(V,N)(T_C-T)^{2-\overline{\alpha }}
\end{equation}
with $\overline{D}(V,N)$ as a parameter and $T_C=T_C(V,N)$. Here $F_0$ can
be interpreted as the Helmholtz potential of the reference system. Repeating
the calculations with the use of (29), quite similar to the presented above
with the use of the Gibbs potential (cf (11)-(17)), we can find the exact
relations similar to (18)-(22)\footnote{%
The same critical relations (30)-(34) can also be obtained with the use of
the exact Helmholtz potential, similarly to systems at constant pressure $p$
(see Appendix A) and therefore the formulae (30)-(34) are exact.}. We obtain 
\begin{equation}
\lim\limits_{T\rightarrow T_C^{-}}\frac{\Delta (\frac{\partial \mu }{%
\partial T})_{V,N}}{\Delta c_{V,N}}=\left( \frac{\partial \ln T_C}{\partial N%
}\right) _V,
\end{equation}

\begin{equation}
\lim\limits_{T \rightarrow T_C^-} \frac{\Delta (\frac{\partial p}{\partial T}%
)_{V,N}}{\Delta c_{V,N}}=-\left( \frac{\partial \ln T_C}{\partial V}
\right)_N,
\end{equation}

\begin{equation}
\lim\limits_{T \rightarrow T_C^-} \frac{\Delta (\frac{\partial \mu}{\partial
V})_{T,N}}{\Delta c_{V,N}}=-\left( \frac{\partial \ln T_C}{\partial N}
\right)_V \left( \frac{\partial T_C}{\partial V} \right)_N,
\end{equation}

\begin{equation}
\lim\limits_{T\rightarrow T_C^{-}}\frac{\Delta (\frac{\partial \mu }{%
\partial N})_{T,V}}{\Delta c_{V,N}}=-\left( \frac{\partial \ln T_C}{\partial
N}\right) _V\left( \frac{\partial T_C}{\partial N}\right) _V,
\end{equation}
and 
\begin{equation}
\lim\limits_{T\rightarrow T_C^{-}}\frac{\Delta (\frac{\partial p}{\partial V}%
)_{T,N}}{\Delta c_{V,N}}=\left( \frac{\partial \ln T_C}{\partial V}\right)
_N\left( \frac{\partial T_C}{\partial V}\right) _N.
\end{equation}
There are again only five exact expressions of this type because of the
Maxwell relations $\left( \frac{\partial S}{\partial V}\right)
_{T,N}=-\left( \frac{\partial p}{\partial T}\right) _{V,N}$, $\left( \frac{%
\partial S}{\partial N}\right) _{T,V}=-\left( \frac{\partial \mu }{\partial T%
}\right) _{V,N}$ and $\left( \frac{\partial p}{\partial N}\right)
_{T,V}=-\left( \frac{\partial \mu }{\partial V}\right) _{T,N}$. There exists
also a freedom to form another quotiens what results in many other
expressions easy to obtain (we don't write them explicitely here). The
analogous cross-relations, similar to (24)-(26) have the form 
%\begin{widetext}
\begin{equation}
\lim\limits_{T\rightarrow T_C^{-}}\frac{\Delta \left( \frac{\partial \mu }{%
\partial V}\right) _{T,N}}{\Delta c_{V,N}}=T_C\big[\lim\limits_{T\rightarrow
T_C^{-}}\frac{\Delta \left( \frac{\partial \mu }{\partial T}\right) _{V,N}}{%
\Delta c_{V,N}}\big] \cdot \big[ \lim\limits_{T\rightarrow T_C^{-}}\frac{%
\Delta \left( \frac{\partial p}{\partial T}\right) _{V,N}}{\Delta c_{V,N}}%
\big],
\end{equation}

\begin{equation}
\lim\limits_{T \rightarrow T_C^-} \frac{\Delta \left( \frac{\partial \mu }{%
\partial N}\right)_{T,V}}{\Delta c_{V,N}}= -T_{C}\big[\lim\limits_{T
\rightarrow T_C^-} \frac{\Delta \left( \frac{\partial \mu }{\partial T}%
\right)_{V,N}}{\Delta c_{V,N}} \big]^2
\end{equation}
and 
\begin{equation}
\lim\limits_{T\rightarrow T_C^{-}}\frac{\Delta \left( \frac{\partial p}{%
\partial V}\right) _{T,N}}{\Delta c_{V,N}}=T_C\big[\lim\limits_{T\rightarrow
T_C^{-}}\frac{\Delta \left( \frac{\partial p}{\partial T}\right) _{V,N}}{%
\Delta c_{V,N}}\big]^2.
\end{equation}
%\end{widetext}
The other cross-relations are also very easy to find. In the case of
superconducting systems we can drop the $\lim\limits_{T\rightarrow T_C^{-}}$
in these expressions and we obtain the formulae (46)-(53) from Ref. [8].
Similar calculations can also be repeated in the vicinity of the critical
point for the case $T>T_C$, using the form of the specific heat for this
case (see (2) and (27)). It leads, however, to the same critical expressions
(18)-(22), (24)-(26) and (30)-(37) where the limit $T\rightarrow T_C^{-}$
should be replaced by $T\rightarrow T_C^{+}$. In other words, we can replace
in all these expressions the limit $T\rightarrow T_C^{-}$ by a more general
form $\lim\limits_{T\rightarrow T_C}$.

\section{Logarithmic anomaly of the specific heat}

It is interesting to see whether the derived critical relations (18)-(22),
(24)-(26) and (30)-(37) are also valid in the case of logarithmic divergence
of the specific heat, when

\begin{equation}
\Delta c_{p,N}(T)\varpropto (\ln (T_C-T))^a
\end{equation}
where $a=\frac 13$ for uniaxial ferroelectrics (see e.g. Ref. [5], p. 384)
or $a=1$ for liquid $He^4$ (see e.g. Refs [1], [9])\footnote{%
According to different authors for liquid $He^4$ the formula (2) can also be
valid for $\alpha ,\alpha ^{\prime }$ very small (cf e.g. [3] and papers
cited therein). Thus, the critial relations, derived in the preceding
Section are authomatically valid in this case.}.\newline
For simplicity, we restrict ourselves to the case $T<T_C$ where
\begin{equation}
\Delta c_{p,N}(T)=A(p,N)[\ln (T_C-T)]^a
\end{equation}
and $A(p,N)$ is a parameter. It is easy to see that the leading term of the
Gibbs potential, deduced from the expressions (6) and (39) should have the
form\footnote{%
The exact form for $\Delta Z$ is given in Appendix B. Similarly to the
preceding Section, the additional terms in $\Delta Z$ does not contribute to
the final relations (18)-(22) which remain exactly the same. Because of
extremally long expressions we present here only the calculations resulting
from the leading term in the formula (40).}
\begin{equation}
Z(T,p,N)=Z_0(T,p,N)-\frac 12D(p,N)(T_C-T)^2[\ln (T_C-T)]^a
\end{equation}
where $D(p,N)$ is a parameter. Applying the formulae (5) and (6) we obtain
\begin{eqnarray}
\Delta c_{p,N}(T)&=&TD(p,N)\{[\ln (T_C-T)]^a \nonumber \\
&&+\frac{3a}2[\ln (T_C-T)]^{a-1}+\frac{a(a-1)}2[\ln (T_C-T)]^{a-2}\}.
\end{eqnarray}
In the vicinity of the critical point the second and third term in the
parenthesis of (41) can completely be neglected ($a=\frac 13,1$) in
comparison to the first, singular term. Thus, we can write
\begin{equation}
\Delta c_{p,N}(T)\approx TD(p,N)[\ln (T_C-T)]^a\approx T_CD(p,N)[\ln
(T_C-T)]^a.
\end{equation}
In other words the formula for the Gibbs potential (40) reproduces the
critical behaviour of the specific heat (39) when we assume that 
\begin{equation}
A(p,N)=T_CD(p,N).
\end{equation}
The chemical potential $\mu =(\frac{\partial Z}{\partial N})_{T,p}$ ($\mu
_0=(\frac{\partial Z_0}{\partial N})_{T,p}$) and the volume of the system $%
V=(\frac{\partial Z}{\partial p})_{T,N}$ ($V_0=(\frac{\partial Z_0}{\partial
p})_{T,N}$) can be found from (40). We obtain

\begin{eqnarray}
\mu &=&\mu _0-\frac 12\left( \frac{\partial D}{\partial N}\right)
_p(T_C-T)^2[\ln (T_C-T)]^a  \nonumber \\
&-&D(T_C-T)\{[\ln (T_C-T)]^a+\frac a2[\ln (T_C-T)]^{a-1}\}\left( \frac{%
\partial T_C}{\partial N}\right) _p
\end{eqnarray}
and 
\begin{eqnarray}
V &=&V_0-\frac 12\left( \frac{\partial D}{\partial p}\right) _N(T_C-T)^2[\ln
(T_C-T)]^a  \nonumber \\
&&-D(T_C-T)\{[\ln (T_C-T)]^a+\frac a2[\ln (T_C-T)]^{a-1}\}\left( \frac{%
\partial T_C}{\partial p}\right) _N
\end{eqnarray}
where we have taken into account that $T_C=T_C(p,N)$. It is easy to see that
according to (44) and (45) we find

\begin{eqnarray}
\Delta \left( \frac{\partial \mu }{\partial T}\right) _{p,N}&=& \left( \frac{%
\partial D}{\partial N}\right) _p(T_C-T)\{[\ln (T_C-T)]^{a} +\frac{a}{2}[%
\ln(T_C-T)]^{a-1}\}  \nonumber \\
&&+\left( \frac{\partial T_C}{\partial N}\right) _p D\{[\ln(T_C-T)]^{a} 
\nonumber \\
&&+\frac{3a}{2}[\ln (T_C-T)]^{a-1}+\frac{a(a-1)}{2}[\ln (T_C-T)]^{a-2}\},
\end{eqnarray}

\begin{eqnarray}
\Delta \left( \frac{\partial V}{\partial T}\right) _{p,N}&= & \left( \frac{%
\partial D}{\partial p}\right) _N(T_C-T)\{[\ln (T_C-T)]^{a} +\frac{a}{2}[%
\ln(T_C-T)]^{a-1}\}  \nonumber \\
&&+\left( \frac{\partial T_C}{\partial p}\right) _N D\{[\ln(T_C-T)]^{a} 
\nonumber \\
&&+\frac{3a}{2}[\ln (T_C-T)]^{a-1} +\frac{a(a-1)}{2}[\ln (T_C-T)]^{a-2}\},
\end{eqnarray}

\begin{eqnarray}
\Delta \left( \frac{\partial \mu }{\partial p}\right) _{T,N}&= & -\frac
12\left( \frac{\partial ^2D}{\partial p\partial N}\right) (T_C-T)^2[\ln
(T_C-T)]^{a}  \nonumber \\
&&-2\left( \frac{\partial D}{\partial N}\right) _p\left( \frac{\partial T_C}{%
\partial p}\right) _N(T_C-T)\{[\ln (T_C-T)]^{a}  \nonumber \\
&&+\frac{a}{2}[\ln(T_C-T)]^{a-1}\}  \nonumber \\
&&-D\left( \frac{\partial T_C}{\partial p}\right) _N\left( \frac{\partial T_C%
}{\partial N}\right) _p\{[\ln (T_C-T)]^{a}  \nonumber \\
&&+\frac{3a}{2}[\ln (T_C-T)]^{a-1}+\frac{a(a-1)}{2}[\ln (T_C-T)]^{a-2}\},
\end{eqnarray}

\begin{eqnarray}
\Delta \left( \frac{\partial \mu }{\partial N}\right) _{T,p} &=&-\frac
12\left( \frac{\partial ^2D}{\partial N^2}\right) _p(T_C-T)^2[\ln (T_C-T)]^a
\nonumber \\
&&-2\left( \frac{\partial D}{\partial N}\right) _p\left( \frac{\partial T_C}{%
\partial N}\right) _p(T_C-T)\{[\ln (T_C-T)]^a  \nonumber \\
&&+\frac a2[\ln (T_C-T)]^{a-1}\}  \nonumber \\
&&-D[\left( \frac{\partial T_C}{\partial N}\right) _p]^2\{[\ln (T_C-T)]^a+%
\frac{3a}2[\ln (T_C-T)]^{a-1}  \nonumber \\
&&+\frac{a(a-1)}2[\ln (T_C-T)]^{a-2}\}
\end{eqnarray}
and 
\begin{eqnarray}
\Delta \left( \frac{\partial V}{\partial p}\right) _{T,N} &=&-\frac 12\left( 
\frac{\partial ^2D}{\partial p^2}\right) _N(T_C-T)^2[\ln (T_C-T)]^a 
\nonumber \\
&&-2\left( \frac{\partial D}{\partial p}\right) _N\left( \frac{\partial T_C}{%
\partial p}\right) _N(T_C-T)\{[\ln (T_C-T)]^a  \nonumber \\
&&+\frac a2[\ln (T_C-T)]^{a-1}\}  \nonumber \\
&&-D[\left( \frac{\partial T_C}{\partial p}\right) _N]^2\{[\ln (T_C-T)]^a+%
\frac{3a}2[\ln (T_C-T)]^{a-1}  \nonumber \\
&&+\frac{a(a-1)}2[\ln (T_C-T)]^{a-2}\}
\end{eqnarray}
where, as before, $\Delta X\equiv X(T)-X_0(T).$ It is easy to see that
dividing (46)-(50) by (39) with the use of (43) and taking the limit $%
T\rightarrow T_C^{-}$ we obtain exactly the same critical relations
(18)-(22) and also (24)-(26). The calculations in the case $T>T_C$ are quite
similar to the case $T<T_C$ and we do not repeat them here. They also lead
to the same results (18)-(22) and (24)-(26) where the limit $T\rightarrow
T_C^{-}$ should be replaced by $T\rightarrow T_C^{+}.$ In the other words,
similarly to the preceding Section, we can simply use the limit $%
\lim\limits_{T\rightarrow T_C}$ in these expressions. The case of the system
at constant volume $V$ can be treated parallel to the lines desribed above
and in the preceding Section. It, however, leads again to exactly the same
critical relations (30)-(37). Thus, we see that in the case of the specific
heat anomally (2) and (38) the same critical relations are valid.

\section{Conclusions}

Using the phenomenological approach to second order phase transitions we
have shown that independent from the type of the considered system the
critical relations (cross-relations) derived in this paper are exactly the
same for all of them. We have shown in this way that quite different systems
possess the same universal critical behaviour of the quotients. It, however,
means that the exact critical relations, derived in this paper, define a new
class of universal, critical relations valid for a very large number of
magnetic, ferroelectric or superconducting systems including also uniaxial
ferroelectrics and liquid $He^4$.

\section{Appendix A}
\newcounter{rown}[equation]
\setcounter{equation}{0}
\renewcommand{\theequation}{A. \arabic{equation}}
Because of the thermodynamic relation

\begin{equation}
\Delta c_{p,N}=T\left( \frac{\partial \Delta Z}{\partial T}\right) _{p,N}
\end{equation}
we can write

\begin{equation}
\Delta S=\int \frac{\Delta c_{p,N}}TdT+C_1(p,N)
\end{equation}
where $C_1(p,N)$ is an arbitrary constant. In the vinicity of the critical
point the condition $\frac{(T_C-T)}{T_C}\ll 1$ is fulfilled and therefore we
can write

\begin{equation}
\frac 1T=\frac 1{T_C-(T_C-T)}=\frac 1{T_C}\cdot \frac 1{1-\frac{(T_C-T)}{T_C}%
}=\frac 1{T_C}\sum\limits_{l=0}^\infty \frac{(T_C-T)^l}{T_C^l}.
\end{equation}
Thus, we obtain (see the formula (3))

\begin{equation}
\begin{array}{ll}
\Delta S & =\frac{A(p,N)}{T_C}\sum\limits_{l=0}^\infty \frac 1{T_C^l}\int
(T_C-T)^{l-\alpha }dT+C_1(p,N) \\
&  \\
& =-\frac{A(p,N)}{T_C}\sum\limits_{l=0}^\infty \frac{(T_C-T)^{l-\alpha +1}}{%
T_C^l(l-\alpha +1)}+C_1(p,N)
\end{array}
\end{equation}
Because $\alpha <1$ the first term in (A.4) is going to zero at $T=T_C.$ At
the critical point $\Delta S=0$ $(S=S_0)$ and therefore the arbitrary
constant $C_1(p,N)$ has to be equal to zero.

Starting from the expression (A.4) we obtain

\begin{equation}
\begin{array}{ll}
\Delta Z & =-\int \Delta SdT+C_2(p,N) \\
&  \\
& =-\frac{A(p,N)}{T_C}\sum\limits_{l=0}^\infty \frac{(T_C-T)^{l-\alpha +2}}{%
T_C^l(l-\alpha +2)(l-\alpha +1)}+C_2(p,N).
\end{array}
\end{equation}
At the critical point, however, the first term in (A.5) vanishes. Because at
the critical point $\Delta Z=0$ $(Z=Z_0)$ the arbitrary constant $C_2(p,N)$
has to be equal to zero. Thus, we can write the exact expression for $\Delta
Z$ in the form

\begin{equation}
\begin{array}{ll}
\Delta Z & =-\frac{A(p,N)}{T_C}\frac{(T_C-T)^{2-\alpha }}{(2-\alpha
)(1-\alpha )} \\
&  \\
& -A(p,N)\sum\limits_{l=1}^\infty \frac{(T_C-T)^{l-\alpha +2}}{%
T_C^{l+1}(l-\alpha +2)(l-\alpha +1)}
\end{array}
\end{equation}
and we see that the first term (leading term) in (A.6) has the form (7) (see
also (9)). This term properly reproduces the critical behaviour of the
specific heat (see formula (8)) and therefore we have neglected in further
considerations (see (11)-(17)) the second term in (A.6). It is, of course,
possible to take into account the full expression (A.6) for $\Delta Z$ but
it is completely irrelevant. The presence of the second term in (A.6)
modifies the relations (11)-(17) producing very long expresions. It has,
however, absolutely no influence on the critical relations (18)-(22) because
the additional terms in (11)-(17) make no contributions in the limit $%
T\rightarrow T_C^{-}$ when forming the quotients (18)-(22).

\section{Appendix B}
\newcounter{rown1}[equation]
\setcounter{equation}{0}
\renewcommand{\theequation}{B. \arabic{equation}}
Starting from the expression (39) and using (A.3) we find

\begin{equation}
\begin{array}{ll}
\Delta S & =\int \frac{\Delta c_{p,N}}TdT+C_1(p,N) \\
&  \\
& =-\frac{A(p,N)}{T_C}\left[ \ln (T_C-T)\right] ^a\sum\limits_{l=0}^\infty
\frac{(T_C-T)^{l+1}}{T_C^l(l+1)}W_l(T_C-T)+C_1(p,N)
\end{array}
\end{equation}
where

\begin{equation}
W_l(T_C-T)=\sum\limits_{m=0}^\infty a_{l,m}\left[ \ln (T_C-T)\right] ^{-m},
\end{equation}

\begin{equation}
\begin{array}{llll}
a_{l,0}=1, & a_{l,1}=-\frac a{l+1}, & ...\text{ }, & a_{l,m}=(-1)^m\frac{%
a(a-1)\cdot ....\cdot (a-m+1)}{(l+1)^m}
\end{array}
\end{equation}
and $C_1(p,N)$ is an arbitrary constant. The first term in (B.1) vanishes
when $T\rightarrow T_C^{-}$ $(a=\frac 13,1)$. Because $\Delta S$ should
vanish at $T=T_C$ $(S=S_0)$ the arbitrary constant $C_1(p,N)$ should also be
zero. Finaly we find that the exact expression for $\Delta Z$ has the form

\begin{equation}
\begin{array}{ll}
\Delta Z & =-\frac{A(p,N)}{T_C}(T_C-T)^2\left[ \ln (T_C-T)\right] ^a\times
\\
&  \\
& \times \sum\limits_{l=0}^\infty \frac{(T_C-T)^l}{T_C^l(l+1)(l+2)}%
\sum\limits_{m=0}^\infty a_{l,m}\left\{ \left[ \ln (T_C-T)\right]
^{-m}\sum\limits_{n=0}^\infty b_{l,m,n}\left[ \ln (T_C-T)\right]
^{-n}\right\} \\
&  \\
& +C_2(p,N)
\end{array}
\end{equation}
where

\begin{equation}
\begin{array}{cccc}
b_{l,m,0}=1, & b_{l,m,1}=-\frac{a-m}{l+2}, & \cdot , &b_{l,m,n}=(-1)^n \frac{a(a-m)(a-m-1)\cdot ....\cdot (a-m-n+1)}{(l+2)^n}
\end{array}
\end{equation}
and $C_2(p,N)$ is an arbitrary constant. Because the first term in (B.4)
vanishes in the limit $T\rightarrow T_C^{-}$ the arbitrary constant $C_2(p,N)
$ should be equal to zero $(\Delta Z=0$ $(Z=Z_0)$ at $T=T_C).$ The
expression (B.4) can be rewritten in the short form to be

\begin{equation}
\Delta Z=-\frac{A(p,N)}{T_C}(T_C-T)^2\left[ \ln (T_C-T)\right] ^a\left\{
1+R(T_C-T)\right\}
\end{equation}
where $R(T_C-T)$ contains the terms with $l,m,n=1,2,...$ (cf (B.3) and
(B.5)). The first (leading) term in (B.6) has the form of the second term in
expression (40) where $D(p,N)$ is related to $A(p,N)$ by the formula (43).
It is, in principle, possible to perform the calculation (formulae
(44)-(50)) starting from the full expression (B.6). It, however, leads to
extremely long formulae and therefore we have presented the results using
only the leading term. The terms, generated by $R(T_C-T)$ in (B.6) produces
additionally extremely long expressions in (44)-(50) which are completely
irrelevant when forming the quotients (18)-(22) and taking the limit $%
T\rightarrow T_C^{-}$ because their contribution to the expressions
(18)-(22) is equal to zero in this limit. Therefore the critical relation
(18)-(22) are also exact in the case of the logarithmic divergence of the
specific heat (39).

\end{document}